\documentclass{jfm}

\usepackage{epsfig}

\providecommand\upi{\pi}
\def\etal{\mbox{\it et al.\ }}
\newcommand{\curl}{\mbox{${\rm curl \,} $}}
\def\Rey{\mbox{\textit Re}}   
\def\bnabla{\bm{\nabla}}
\def\bcdot{\bm{\cdot}}
\newcommand{\bm}[1]{\mbox{\boldmath $#1$}}
\newcommand{\ov}[2]{\mbox{${\frac{#1}{#2}}$}}

\def\vn{{\bm v}^n}
\def\vs{{\bm v}^s}

\def\v0{{\bm v}^0}
\def\pn{p^n}
\def\ps{p^s}
\def\psin{\psi^n}
\def\psis{\psi^s}
\def\rhon{\rho^n}
\def\rhos{\rho^s}
\def\nun{\nu^n}
\def\nus{\nu^s}
\def\omegahat{\widehat{\bm{\omega}}^s}

\newsavebox{\omegabox}
\sbox{\omegabox}{\boldmath$\omega$}
\newcommand{\omegabold}{\usebox{\omegabox}}
\newsavebox{\phibox}
\sbox{\phibox}{\boldmath$\phi$}

\def\omegas{\omegabold^s}

\title[Transition to Taylor vortex flow in superfluid helium]
{Transition from Ekman flow to Taylor vortex flow 
in superfluid helium}

\normalsize
\author[K. L. Henderson and C. F. Barenghi]%
{K\ls A\ls R\ls E\ls N\ns L.\ns H\ls E\ls N\ls D\ls E\ls R\ls S\ls O\ls N$^1$\ns
\and \break C\ls A\ls R\ls L\ls O\ns F.\ns B\ls A\ls R\ls E\ls N\ls G\ls H\ls I$^2$}

\affiliation{$^1$School of Mathematical Sciences, University of the
West of England, \break Bristol, BS16 1QY, UK. \break
e-mail: Karen.Henderson@uwe.ac.uk\\[\affilskip]
$^2$School of Mathematics, University of Newcastle, \break
Newcastle upon Tyne, NE1 7RU, UK. \break
e-mail: C.F.Barenghi@ncl.ac.uk}

\begin{document}

\maketitle

\begin{abstract}
By numerically computing the steady axisymmetric flow of helium~II confined
inside a finite aspect ratio
Couette annulus, we determine the transition from Ekman flow to
Taylor vortex flow as a function of temperature and aspect ratio. 
We find that the low-Reynolds number flow is quite different
to that of a classical fluid, particularly at lower temperatures.
At high aspect ratio our results confirm
the existing linear stability theory of the onset of Taylor vortices,
which assumes infinitely long cylinders.
\end{abstract}

\section{The equations of motion of superfluid helium}

As the temperature is reduced through $T=T_\lambda=2.1768$~K,
liquid helium undergoes a phase transition from helium~I (a classical
Navier--Stokes fluid) to helium~II (a superfluid). 
The superfluid state persists through to absolute zero at vapour pressure. 
The motion of helium~II is well described by the hydrodynamic two-fluid
theory of Landau \& Tisza (1987).  According to this theory,
helium~II comprises two perfectly mixed fluids,
the viscous normal fluid and the inviscid superfluid, of densities
$\rhon$ and $\rhos$ respectively.  The total density of helium~II,
$\rho=\rhon+\rhos$, does not vary much with the temperature $T$,
whereas $\rhon$ and $\rhos$ depend strongly on $T$.  At absolute
zero, $T=0$~K, the normal fluid component vanishes and helium~II is entirely
superfluid $(\rhon=0)$, whilst at the lambda point, $T=T_\lambda$, the
superfluid component is zero $(\rhos=0)$ and helium~II becomes helium~I,
which is a classical Navier--Stokes fluid.
If helium~II is rotated with angular velocity $\Omega$ greater than some 
small critical value, then vortex filaments appear in the superfluid 
(Donnelly, 1991).  The circulation around each
vortex filament is quantised, in that
\begin{equation}
\oint_C \vs \bm{ \cdot dl}=\Gamma
\end{equation}
where $\vs$ is the superfluid velocity field,
$\Gamma=9.97\times 10^{-4}~$cm$^2\,$sec$^{-1}$
is the quantum of circulation (the ratio of Plank's constant and the
mass of one helium atom) and $C$ is an arbitrary integration
path around the axis of the filament.
When helium~II rotates the vortices align themselves to the direction of 
rotation and form a regular configuration with
areal density (number of vortex filaments crossing the unit area perpendicular
to the direction of rotation)
$N=2\Omega/\Gamma$ (Feynman's rule). 

The hydrodynamics of the superfluid state
is an interesting topic {\it per se},
but it is worth mentioning that attention to this problem has
additional motivations.
The first arises from the engineering applications:
helium is the only substance available in
liquid form at temperatures near absolute zero, so it is important
as a cryogenics coolant. Applications range from infrared detectors
in space science to the cooling of superconducting magnets in
particle physics. The second motivation comes from recent
experimental developments in which the relation between classical and
quantum turbulence is investigated (for example see
Smith \etal 1993; Barenghi, Swanson \& Donnelly 1995).

The most generally accepted equations for modelling the macroscopic
flow of helium~II are the Hall-Vinen-Bekharevich-Khalatnikov (HVBK)
equations which were derived by a number of people over the
years (Hall \& Vinen 1956; Hall 1960; Bekharevich \& Khalatnikov 1965; 
Hills \& Robert 1977).
These equations extend Landau's two-fluid model
to take into account the presence of quantised vortex lines in the flow.
The derivation of the equations is based on a continuum approximation,
assuming a high density of vortex lines, all aligned roughly in the same
direction.  
For such situations, the superfluid vorticity, which is discrete in
nature, may be approximated as a continuum, resulting in an
effective superfluid
vorticity field $\omegas=\curl \vs$.
The isothermal, incompressible HVBK equations may be written as:
\begin{equation}
 \frac{\partial \vn}{\partial t}
+ (\vn \bm{\cdot\nabla }) \vn
= - \bnabla \pn
+ \nun \nabla^2 \vn
+ \frac{\rhos}{\rho} {\bm F},
\label{eq:vneqn}
\end{equation}
\begin{equation}
 \frac{\partial \vs}{\partial t}
+ (\vs \bm{\cdot\nabla }) \vs
= - \bnabla \ps +
{\bm T} - \frac{\rhon}{\rho} {\bm F},
\label{eq:vseqn}
\end{equation}
\begin{equation}
\bnabla \bcdot \, \vn=0, \quad
\bnabla \bcdot \, \vs=0,
\label{eq:div}
\end{equation}
\noindent
where $\vn$ is the normal fluid velocity, $\nun$ is the kinematic 
viscosity of the normal fluid and $\pn$, $\ps$ are effective pressures 
($\bnabla \ps=(1/\rho)\bnabla p -\ov{1}{2}(\rhon/\rho)\bnabla (\vn-\vs)^2$
and $\bnabla \pn=(1/\rho)\bnabla p+\ov{1}{2}(\rhos/\rho)\bnabla (\vn-\vs)^2$
where $p$ is the pressure).
The mutual friction force may be written as
\begin{equation}
{\bm F}= \ov{1}{2}B \omegahat \!\times
(\omegas \times (\vn-\vs-\nus\bnabla\times\omegahat)) +
\ov{1}{2}B'\omegas\! \times 
(\vn-\vs-\nus\bnabla\times\omegahat)
\label{eq:MF}
\end{equation}
with $\omegahat=\omegas/|\omegas|$ the unit
vector in the direction of superfluid vorticity and $B$, $B'$ are
the temperature-dependent mutual friction
parameters  (Barenghi, Donnelly \& Vinen 1983, Donnelly \& Barenghi 1998).
This force is due to collisions between the normal fluid (mainly rotons,
at temperatures relevant to most experiments) and vortex lines.
The vortex tension force may be written as
\begin{equation}
 \bm{T}= -\nus \omegas\times (\bnabla \!\times \!\omegahat),
\label{eq:VT}
\end{equation}
and reflects the energy per unit length in the vortex lines.
The vortex tension parameter
\begin{equation}
\nus=(\Gamma / 4\upi) \log (b_{_0}/a_{_0})
\end{equation}
has the same dimension as kinematic viscosity but physically it
is very different: it represents the ability of a vortex line to
oscillate due to vortex waves which can be excited on the vortex
lines themselves.
The quantity $b_{_0}=(|\omegas|/\Gamma)^{-1/2}$ represents the
intervortex spacing and $a_{_0}\approx 10^{-8}$ cm is the radius of 
the superfluid vortex core.

\section{Motivation and aim of this work}

\begin{figure}
\vspace{70mm}
\includegraphics{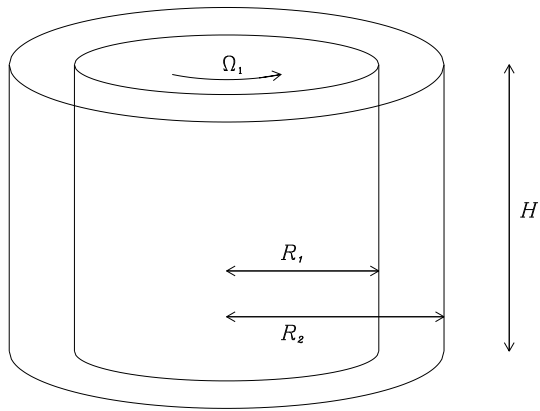}
\caption{The problem's configuration. The fluid is contained
inside a cylindrical box of inner radius $R_1$, outer radius
$R_2$ and height $H$.
The inner cylinder rotates at constant
angular velocity $\Omega_1$, while the outer cylinder and the
top and bottom end plate are stationary.}
\label{fig:couette}
\end{figure}

In this paper we apply the HVBK equations to model helium~II in a 
finite aspect ratio Couette annulus, that is flow of fluid confined radially
between two concentric rotating cylinders and axially between two
fixed plates separated by a distance $H$.   We consider
the inner cylinder ($r=R_1$) to be rotating with angular velocity $\Omega_1$
and the outer cylinder ($r=R_2$) to be fixed, see figure~\ref{fig:couette}.
Taylor--Couette flow has been used as a bench-mark for fluid mechanics
since Taylor's pioneering work (1923) to investigate the
transition from Couette flow to Taylor vortices, which established a
firm ground for using the Navier--Stokes equations and the no-slip
boundary conditions.  Progress in helium~II has been slower than for
classical fluids due in part to problems of flow visualisation at such
low temperatures.  In considering a
classical fluid, introduction of flakes or other small particles into
the working fluid (usually oil or water), results in the Taylor vortices
being clearly evident.  In contrast there are only limited visualisation
techniques available to the experimentalist at temperatures close to
absolute zero.
Recent attempts have been made to reveal the
flow pattern of helium~II by adding small particles (Bielert \& Stamm 1993).
However this was only successful at high rotation rates (40 times the
critical angular velocity at which linear stability analysis predicts
Couette flow becomes unstable); in this regime the flow is 
turbulent and the validity of the HVBK equations is not clear.
Clearly the lack of direct flow visualisation gives additional motivation 
to our work.

Experiments on helium~II between concentric cylinders were first
performed by Kapitza in 1941 and
Donnelly \& LaMar (1988) have written a review of
experiments involving helium~II in Couette apparatus.  
Early Taylor--Couette experiments
were concerned with determining the viscosity of helium~II by
measuring the torque exerted by the flow on the stationary cylinder.
A break in the linear dependence of the torque with the angular
velocity of the rotating cylinder is taken to denote a transition
from one solution to another.
A second experimental technique is available, that of measuring the extra
attenuation of a second sound wave, which can be used to probe the
superfluid vorticity.
Second sound waves occur when there is a
periodic counterflow between the normal fluid and superfluid, which
corresponds to a wave of heat.  Angular velocity is plotted against
the attenuation factor and breaks in the curve are interpreted as
transitions in the flow.  By measuring the
extra attenuation of second sound waves in the axial, azimuthal and
radial directions it is theoretically possible to get an idea of the
number and direction of the quantised vortex lines.  In practice the
information obtained is less complete than this.

Despite the fact that the HVBK equations were established in 1961, they
have been used mainly to model helium~II in a rotating
container (solid body rotation).
Unfortunately solid body rotation (which corresponds to a spatially uniform
configuration of superfluid vortices) is not a strict test of
the equations of motion because it is too simple; the terms in
the HVBK equations which involve the vortex tension vanish.
Only recently have the HVBK equations
been validated for a non-trivial flow configuration: Couette flow.
For the case of infinitely long cylinders, both superfluid and normal
fluid velocity fields have the following Couette flow profile:
\begin{equation}
 \vn=\vs=\frac{(-\Omega_1 R_1^2\,r+
\Omega_1 R_1^2R_2^2\,r^{-1})}{(R^2_2-R^2_1)} \widehat{\bm{\phi}}
\label{eq:couette}
\end{equation}
where $(r,\phi,z)$ are cylindrical coordinates and
$\widehat{\bm{\phi}}$ is the unit vector in the azimuthal  direction.
After making the usual assumption of infinitely long cylinders,
Barenghi \& Jones (1988) and Barenghi (1992)
considered the linear stability of Couette flow with respect to
infinitesimal perturbations of the form $\exp({\rm i}m\phi+{\rm i}kz+\sigma t)$
where $\sigma$ is the complex growth rate, $m$ the azimuthal wavenumber
and $k$ the dimensionless axial wavenumber
(expressed in units of $1/\delta$ where $\delta=R_2-R_1$ is the gap width).
The driving parameter of the problem is the Reynolds number,
\begin{equation}
\Rey=\frac{\Omega_1 R_1 \delta}{\nun},
\label{eq:re}
\end{equation}
which represents the dimensionless velocity of the inner cylinder.
Barenghi \& Jones (1988) found that if the inner cylinder rotates
sufficiently fast, at a certain critical value $\Rey=\Rey_{\rm{crit}}$
the growth rate, Real$(\sigma)$ of the axisymmetric ($m=0$)
perturbation becomes positive, hence Couette flow becomes unstable. This
transition corresponds to the onset of Taylor vortices for a classical
fluid. Barenghi \& Jones (1988) also determined the temperature dependence
of the critical Reynolds number, $\Rey_{\rm{crit}}$, and of the critical
axial wavenumber, $k_{\rm{crit}}$.
Their work prompted further experiments and good agreement between 
the predicted and measured values of $\Rey_{\rm{crit}}$ was
found (Barenghi 1992), particularly for temperatures close to the
lambda point.  The success of the linear stability analysis
prompted further work, namely the study of nonlinear
Taylor flow in infinitely long cylinders (Henderson, Barenghi \& 
Jones 1995).
Comparisons with existing experimental data and numerical results further
validated the HVBK model in the high temperature regime (Henderson \& 
Barenghi 1994).

Unfortunately at lower temperatures ($T<2$~K), although there was
qualitative agreement, the measured critical Reynolds numbers were
larger than those predicted by the linear stability analysis. 
In the classical Taylor--Couette
problem the critical value of the dimensionless axial wavenumber
is $k_{crit}\approx \upi$, hence each individual Taylor vortex cell
is approximately square (the extension in the axial direction is 
equal to the gap's size). In the case of helium~II, Barenghi \& Jones
(1998) found that $k_{crit} \to \upi$ as $T\to T_{\lambda}$, as expected
in the limit of a pure normal fluid. However,
as the temperature is reduced, they found that
$k_{crit}$ decreases and
tends to zero in the limit of a pure superfluid. This result suggests
that the discrepancy between theory and experiments for $T<2$~K is due
to end effects:  there are not enough Taylor vortex cells in
the typical experimental apparatus, thus the infinite cylinder assumption 
breaks down. A better theory is required and that is what we set out
to do in this paper.

It is known in the classical
Taylor--Couette literature that the presence of fixed ends induces
a large scale Ekman circulation in which the fluid moves radially
inward near the top and bottom ends of the Couette apparatus 
and moves radially outward in the middle. This effect is caused by the
no-slip boundary conditions; the centrifugal force pushes the fluid
outwards at the centreline, where the braking effect of the end plates
is least, so the fluid near the end plates moves inwards to conserve
mass.  In the case of helium~II the normal fluid
obeys the same no-slip boundary condition of an ordinary classical fluid,
but there are also superfluid boundary conditions to take into
account.  A discussion of what these superfluid boundary conditions
should be is contained in our previous paper (Henderson \& Barenghi 2000), 
in which we also showed that the competition between the normal fluid and
superfluid boundary conditions has unexpected effects on the
direction of rotation of the Ekman circulation in a unit aspect ratio
annulus. The aim of the current paper is to understand how the appearance
of Taylor vortex flow is effected by the underlying Ekman circulation. 
Firstly we shall investigate the low Reynolds
number flow of helium~II at varying aspect ratios to see how it 
differs from the flow of a classical Navier--Stokes fluid.  Secondly,
we shall determine the transition from Ekman cells to Taylor vortices
and compare the Taylor--Couette flow of helium~II in an 
enclosed annulus to the flow in an infinite apparatus (Henderson \etal 1995).
If, for sufficiently long
cylinders, the results of the linear stability theory are recovered,
contact will be made between theory and experiments at low temperatures.

\section{Model}

We consider the fluid to be confined radially
between two concentric cylinders of inner and outer radius
$R_1$ and $R_2$,
and axially between two fixed plates which are separated by a
distance $H$.  The top and bottom plates and the outer
cylinder are held stationary  and the inner cylinder rotates at
constant angular velocity $\Omega_1$.  Throughout this work we shall
consider the radius ratio, $\eta=R_1/R_2=0.976$ as in the experimental
apparatus of Swanson \& Donnelly (1991) and will vary the
Reynolds number, $\Rey=\Omega_1 R_1 \delta/\nun$, the aspect ratio,
$h=H/\delta$ and the temperature, $T$.

In order to solve the HVBK equations~(\ref{eq:vneqn}-\ref{eq:div})
we need boundary conditions.
The boundary conditions for the normal fluid are no-slip, that is,
working in cylindrical coordinates
$\vn=\Omega_1 R_1 \widehat{\bm{\phi}\,}$
at $r=R_1$ and
$\vn=0$ at $r=R_2,$ $z=0, z=H$.
The boundary conditions for the superfluid are more delicate.  Given
$\bm{n}$, a normal to the boundary, we
have that $\vs \bcdot \, \bm{n}=0$ at $r=R_1, R_2$ and $z=0,H$,
ensuring no
flow normal to the boundaries.  For the remaining conditions
we have taken
\begin{equation}
 \omegas \times \widehat{\bm{z}}=0
\hspace{3mm} {\rm at} \hspace{3mm} r=R_1, R_2 \hspace{3mm} {\rm and}
\hspace{3mm} z=0, H,
\end{equation}
where $\widehat{\bm{z}}$ is the unit vector in the axial direction.
Thus the superfluid vorticity is taken to be
purely axial at the boundaries.  The
condition on the cylinder walls, $r=R_1$ and $r=R_2$, has been discussed
in a previous paper (Henderson \etal 1995) whilst the condition on the
ends of the cylinder, $z=0$ and $z=H$, corresponds to perfect sliding of
the vortex lines (Khalatnikov 1965).  It must be stressed that as yet there
is no experimental evidence for the boundary conditions which we propose;
in fact, a motivation behind or work is to explore the consequence of
assuming certain boundary conditions, hoping to stimulate experimental
work on this issue.  Other possibilities are that the
vortex lines remain totally or partially pinned at the boundaries.
Total pinning of the vortex lines on the ends of the cylinders cannot occur
because each rotation would wrap up the vortex lines until de-pinning 
takes place.  Partial sliding is a possibility, however this would 
introduce extra unknown parameters into the problem depending on the nature 
and the smoothness of the boundaries, so we have chosen to adopt perfect
sliding.

We make the simplifying assumption that the flow is axisymmetric and 
introduce stream functions $\psin$, $\psis$ such that
\begin{subeqnarray}
\gdef\thesubequation{\theequation\textit{a,b}}
v^p_r&&=-\frac{1}{r}\frac{\partial \psi^p}{\partial z}, \hspace{8mm}
v^p_z=\frac{1}{r}\frac{\partial \psi^p}{\partial r}
\label{eq:psi}
\end{subeqnarray}
\returnthesubequation
where $p=n$, $s$ represents the normal fluid, superfluid respectively.
Introducing the stream functions 
ensures that continuity~(\ref{eq:div}) is automatically satisfied.
A finite difference approach is used to obtain solutions for $\psi^p$, 
$v^p_\phi$ and $\omega^p_\phi$.  The equations satisfied by these 
quantities are obtained by taking the $\phi$ component of
(\ref{eq:vneqn},\ref{eq:vseqn}) and the $\phi$ component of
the curl of (\ref{eq:vneqn},\ref{eq:vseqn}).  A Poisson's equation
links the stream function $\psi^p$ with the azimuthal component of
vorticity $\omega^p_\phi$. These equations are solved on the 
computational domain $R_1 \le r \le R_2$, $0 \le z \le H/2$ and symmetry
is assumed to calculate the derivatives on the centreline $z=H/2$.
A uniform grid is used in both the
radial and axial direction and the equations are stepped forward in 
time until a steady solution is achieved. The computational cost given
our criterion of convergence determines the largest aspect ratio
which we compute ($h=8$). 
The equations are made dimensionless using the gap width $\delta$ as 
the unit of length and the normal fluid viscous time scale $\delta^2/\nun$
as the unit of time.
For further details of the numerical method, see Henderson \&
Barenghi (2000).

\section{Results}

In this section we refer to previous results~(Henderson \&
Barenghi 2000) in which
we considered the low Reynolds number flow of helium~II in a unit 
aspect ratio Couette annulus; this geometry constrains the flow to a simple
Ekman circulation.  The key discovery was the 
anomalous Ekman motion of helium~II, compared to a classical Navier--Stokes
fluid.  We found that the pair of superfluid Ekman cells always
rotate in a counter-classical direction (which means for
example that $v^s_r<0$ at the centreline
$z=H/2$), whilst the normal fluid Ekman
cells rotate classically ($v^n_r>0$ at $z=H/2$) 
at temperatures close to the lambda
temperature, but reverse ($v^n_r<0$ at $z=H/2$)
at lower temperatures.  It was also found that 
the azimuthal superfluid velocity is almost independent of $z$.  This
effect becomes more pronounced at lower temperatures and was found
to be due to the tension in the superfluid vortex lines.

\begin{figure}
\vspace{70mm}
\includegraphics{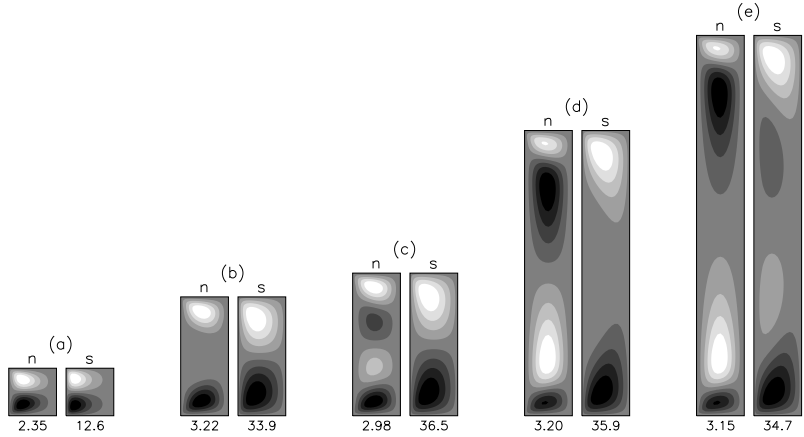}
\caption{Contour plots of the stream function of the normal fluid (n)
and the superfluid (s) at $T=2.11$~K and $\Rey=100$ at aspect ratios:
(a) $h=1$, (b) $h=2.5$, (c) $h=3$, (d) $h=6$, (e) $h=8$.}
\label{fig:fig2}
\end{figure}

\begin{figure}
\vspace{50mm}
\includegraphics{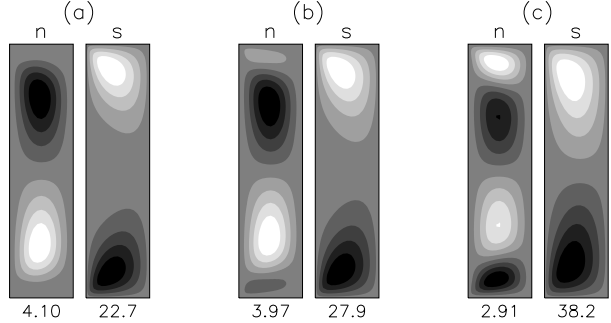}
\caption{Contour plots of the stream function of the normal fluid (n)
and the superfluid (s) at $\Rey=100$ and $h=4$ at temperatures:
(a) $T=2.16$~K, (b) $T=2.15$~K, (c) $T=2.1$~K.}
\label{fig:fig3}
\end{figure}

\begin{figure}
\vspace{55mm}
\includegraphics{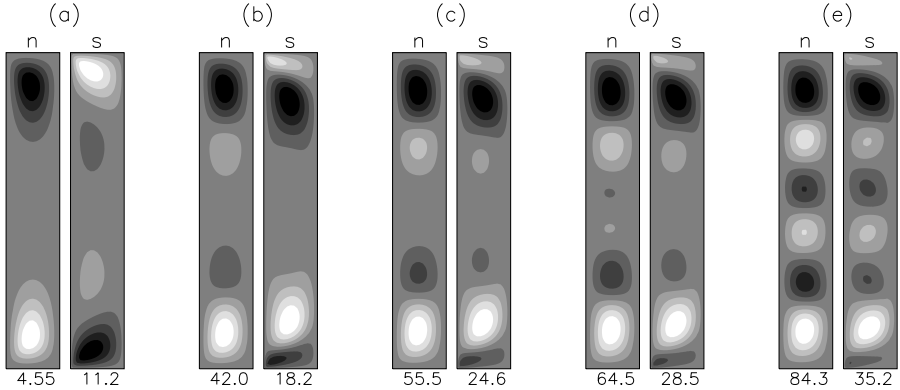}
\caption{Contour plots of the stream function of the normal fluid (n)
and the superfluid (s) at $T=2.17$~K and $h=6$ at Reynolds numbers:
(a) $\Rey=100$, (b) $\Rey=250$, (c) $\Rey=270$, (d) $\Rey=280$,
(e) $\Rey=300$.}
\label{fig:fig4}
\end{figure}

In this paper, the parameters which are varied are the temperature 
$T$, Reynolds number $\Rey$ of the inner cylinder and aspect ratio $h$.
Figures~\ref{fig:fig2}--\ref{fig:fig4},\ref{fig:fig6} 
show contour plots of the normal and superfluid stream functions, where 
the plots are shown for the complete cross-section of the annulus;
$R_1 \le r \le R_2$, $0\le z \le H$
with the inner/outer cylinder on the left/right respectively.  The 
maximum value of each field is printed underneath the corresponding contour
plot. Light and dark regions correspond to positive and negative contour
lines respectively.

In figure~\ref{fig:fig2} we investigate how varying the aspect ratio 
effects the flow of helium~II at $T=2.11$~K and $\Rey=100$.  
This Reynolds number is appreciably below the critical Reynolds number 
$(\Rey_{\rm crit}=355)$ at which the
linear stability analysis predicts Couette flow becomes unstable in
the infinite cylinder approximation. It is apparent from 
figure~\ref{fig:fig2}\,(\textit{a})
that since the normal fluid and superfluid stream functions increase 
with $z$ across the centreline $z=H/2$, (\ref{eq:psi}\,\textit{a}) predicts
that the radial velocity components are negative at the centreline,
in contrast to the motion of a classical Navier--Stokes fluid.
Thus for small aspect ratios ($h\le 2.5$) both the normal fluid and 
superfluid rotate
counter-classically in a pair of matching Ekman cells.  This behaviour has
been reported previously~(Henderson \& Barenghi 2000) and is due 
to the boundary 
conditions satisfied by the vortex lines on the top and bottom of the 
cylinders.  
Figure~\ref{fig:fig2} also shows that the superfluid Ekman cells
fill the whole annulus, becoming more elongated as the aspect ratio is 
increased.  In contrast, the normal fluid Ekman cells, tend to retain their
size and are positioned towards the ends of the cylinders.

As the aspect ratio is increased further ($h> 2.5$)
the superfluid remains in the form of two 
counter-rotating cells whilst the normal fluid splits 
into four cells with the cells adjacent to the ends of the cylinder 
rotating in the same direction as the superfluid.  
The inner cells of the normal fluid strengthen and fill the annulus
as the aspect ratio is increased.  At  aspect 
ratio $h=8$ the superfluid splits into four cells and although the size
of the cells is different to those of the normal fluid, the direction 
of rotation is the same for both fluids.  

Taken together these results demonstrate that 
the low-Reynolds number flow of helium~II is dominated by end effects,
which force the outer cells of the normal and superfluid to rotate
counter-classically.  As the aspect ratio is increased the influence of 
the ends at the centre of the apparatus becomes weaker and classical 
behaviour is observed in both fluids, with outflow at the centreline,
$z=H/2$.  In our previous work (Henderson \&
Barenghi 2000) we highlighted the columnar motion of the azimuthal 
component of superfluid velocity.  We find that as the aspect ratio is 
increased this column like motion becomes less pronounced.

In figure~\ref{fig:fig3} we show  how varying the temperature 
effects the flow of helium~II at $\Rey=100$ and $h=4$.  At this aspect 
ratio, the maximum axial wavelength of a disturbance is, 
$\lambda_{\rm max}=4$ giving a corresponding minimum admissible axial 
wavenumber of $k_{\rm min}=\upi/2$.  Given this, the Reynolds number, 
$\Rey=100$ is below the critical value predicted by linear stability 
analysis at which Couette flow become unstable for all temperatures
considered.  At temperatures close to the lambda temperature ($T\ge 2.16$K)
the normal fluid and superfluid rotate in a pair of Ekman cells.  The 
normal fluid rotates in a classical direction and the superfluid rotates 
counter-classically.  As the temperature is reduced the normal 
fluid splits into four cells, whilst the superfluid remains as two cells.
In this region, the normal fluid cells closest to the ends 
of the cylinder rotate in the same direction as the 
superfluid and become stronger in magnitude and also fill 
more of the annulus as the temperature is reduced.  
The columnar behaviour of 
the azimuthal superfluid velocity is not evident at temperatures close to 
the lambda temperature but becomes more pronounced as the temperature 
is reduced.

\begin{figure}
\vspace{70mm}
\includegraphics{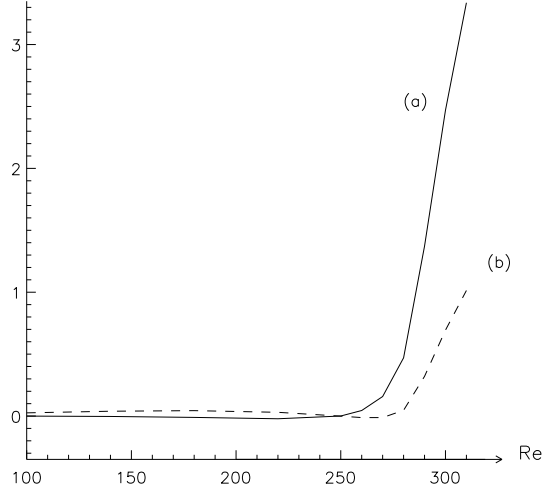}
\caption{Plot of the average values of (a) $v_r^n$ (solid) and (b) $v_r^s$
(dashed) at the centreline $z=H/2$ against Reynolds number.}
\label{fig:fig5}
\end{figure}

\begin{figure}
\vspace{55mm}
\includegraphics{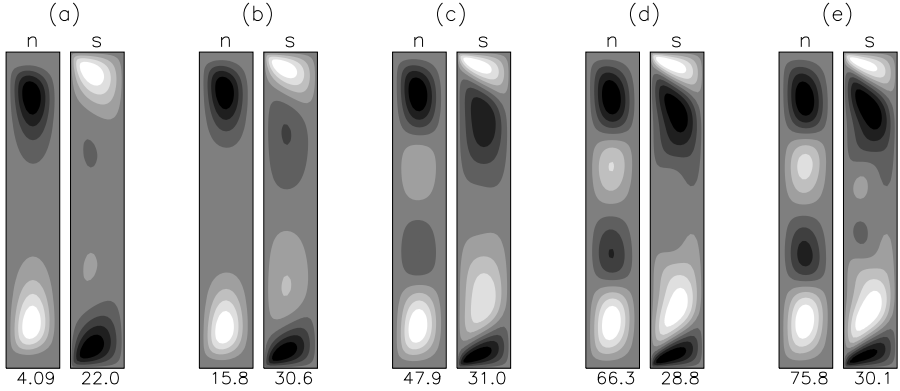}
\caption{Contour plots of the stream function of the normal fluid (n)
and the superfluid (s) at $T=2.16$~K and $h=6$ at Reynolds numbers:
(a) $\Rey=100$, (b) $\Rey=180$, (c) $\Rey=275$, (d) $\Rey=300$,
(e) $\Rey=310$.}
\label{fig:fig6}
\end{figure}

In figure~\ref{fig:fig4} we investigate the transition from Ekman flow to
Taylor vortex flow
at $T=2.17$~K and $h=6$.  At low
Reynolds numbers ($\Rey=100$) the normal fluid consists of
a pair of classically 
rotating Ekman cells.  The superfluid has four cells, with the cells 
adjacent to the ends of the cylinders rotating counter-classically, as 
predicted by the boundary conditions.  As the Reynolds number is increased 
both the normal fluid and the superfluid develop more cells, with the 
cells closest to the ends rotating classically and counter-classically 
for the normal and superfluid respectively, as a consequence of the
boundary conditions.  At $\Rey=300$  Taylor vortex flow
is fully developed and we can see that despite the superfluid 
having an additional weak pair of cells close to the ends of the cylinders,
the two fluids appear to match each other.  In figure~\ref{fig:fig5}
we plot the centreline average values of the radial components of the 
normal fluid (solid) and superfluid (dashed) against Reynolds number.  From
the graph we can see that the transition to Taylor vortices occurs at
approximately $\Rey=275$, whilst stability analysis in the infinitely
long cylinders approximation predicts 
$\Rey_{\rm crit}=278.7$ at $k_{\rm crit}=2.9$.

In figure~\ref{fig:fig6} we repeat the calculation 
at $h=6$ but setting a lower temperature, $T=2.16$~K.  At low
Reynolds numbers ($\Rey=100$) the normal fluid has a pair of classically
rotating Ekman cells.  As before, the superfluid has four cells, 
and the cells which are
adjacent to the ends of the cylinders rotate counter-classically, as
predicted by the boundary conditions.  As the Reynolds number is increased,
the normal fluid develops four
cells which fill the whole apparatus, not six as at $T=2.17$~K.
The outer cells rotate classically
due to the boundary conditions.  The superfluid remains as four cells up
to $\Rey=300$, with the outer cells rotating counter-classically and the 
inner cells strengthing with increasing Reynolds number.  At $\Rey=310$
the superfluid develops an additional small weak pair of cells at the 
centre of the apparatus.  

\begin{figure}
\vspace{70mm}
\includegraphics{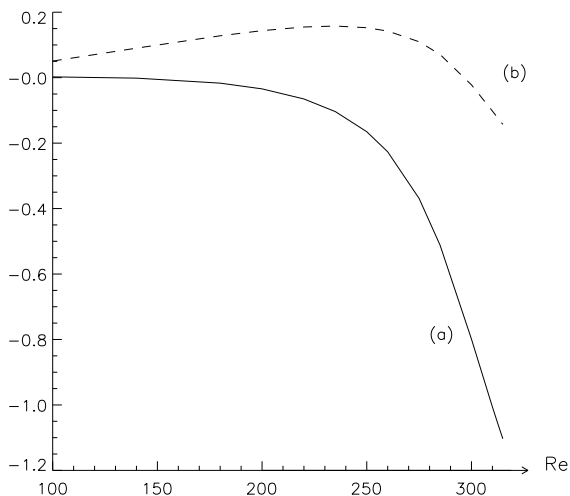}
\caption{Plot of the average values of (a) $v_r^n$ (solid) and (b) $v_r^s$
(dashed) at the centreline $z=H/2$ against Reynolds number.}
\label{fig:fig7}
\end{figure}

In figure~\ref{fig:fig7}
we plot the centreline average values of the radial components of the
normal fluid (solid) and superfluid (dashed) against Reynolds number.  
The transition from the Ekman flow to the  Taylor vortex flow is
harder to distinguish than at higher temperatures and is slightly 
different for the normal and superfluid.  It occurs at approximately 
$\Rey=270$.  What is interesting about these results is that
the Taylor vortices for the normal fluid are elongated, being $1\ov{1}{2}$
times longer than they are wide.  This is consistent with the linear 
stability analysis which predicts that $\Rey_{\rm crit}=310.68$ at the 
critical axial wavenumber of $k_{\rm crit}=2.35$.  The elongation of the
Taylor cells was first predicted by Barenghi \& Jones (1988)
and this is the first time that it has been observed numerically taking 
end effects into account.

\begin{figure}
\vspace{55mm}
\includegraphics{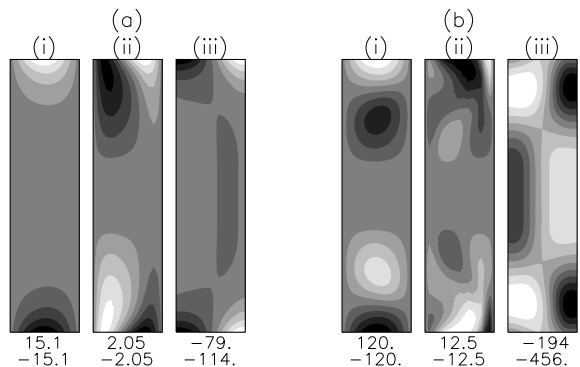}
\caption{Contour plots
of the components of superfluid vorticity:
(i) $\omega^s_r$, (ii) $\omega^s_\phi$, (iii) $\omega^s_z$ at $T=2.16$~K,
$h=6$ for (a) $\Rey=100$, (b) $\Rey=300$
in the middle two-thirds
of the annulus, $1\le z/\delta\le 5$.  The maximum and minimum values of
each field are printed underneath the corresponding plot. }
\label{fig:fig8}
\end{figure}

We find that in general the superfluid vorticity is primarily 
axial and concentrated towards the ends of the cylinders close to the 
inner cylinder.  This predicted result could be observed 
experimentally by measuring the extra attenuation of second sound waves 
due to the vortex lines at the centre and ends of the apparatus.  
As the Reynolds number is increased, there is a much larger
deflection of the vortex lines in the radial direction.  This deflection
has been observed experimentally in the transition to Taylor 
vortices~(Swanson \& Donnelly 1991).  In figure~\ref{fig:fig8} we plot the 
superfluid vorticity components of helium~II at $T=2.16$~K, $h=6$ 
at the Reynolds numbers (a) $\Rey=100$ and (b) $\Rey=300$ for the middle
two-thirds of the apparatus ($1\le z/\delta\le 5$).  
This enables us to see what 
happens to the vortex lines in the centre of the apparatus away from
the cylinder ends, where there is stronger shear which would make the
contour lines in the middle of the apparatus less visible.  
The deflection in the radial direction is clearly
visible for $\Rey=300$.  In the Taylor vortex flow regime, the vorticity 
pattern is similar to that found when considering the Taylor--Couette flow
of helium~II without end effects~(Henderson \etal 1995).

\section{Conclusions} 

By solving the nonlinear HVBK equations in a finite aspect ratio 
configuration we have determined how the fixed ends of the cylinders
effect the Ekman flow and the transition to Taylor vortex flow.

For low-Reynolds number flow, the main result is the anomalous Ekman
circulation of
helium~II when compared to that of a classical Navier--Stokes fluid.  At
short aspect ratios both normal fluid and superfluid rotate as a pair
of Ekman cells.  The superfluid cells always rotate counter-classically,
whilst the sense of rotation of the normal fluid is temperature dependent.
At larger aspect ratios the influence of the ends of the cylinders 
towards the centre of the apparatus diminishes and at lower temperatures
the normal fluid develops additional cells resulting in classical outflow
at the centreline.  The superfluid also develops extra cells to match the 
flow of the normal fluid at the centre.

We have found that at
$T=2.17$~K and aspect ratio as small as $h=6$, in the regions way from the
ends of the cylinders, the superfluid's pattern matches that of 
the normal fluid.
The transition to Taylor vortex flow occurs at a Reynolds number close to 
that predicted by the linear stability analysis in the infinite cylinder
approximation.  At the slightly lower temperature of
$T=2.16$~K we find that the normal fluid develops elongated Taylor cells,
as predicted by the linear stability analysis.  The superfluid flow pattern
is similar to that of the normal fluid away from the ends of the cylinders
but the matching is not as pronounced as at higher temperatures.

Finally we note that gaining more insight into the flow of helium II is 
particularly worthwhile because of the lack of direct flow visualisation
at temperature close to absolute zero.

\bibliographystyle{jfm}

\end{document}